# The Effects of the Introduction of Bitcoin Futures on the Volatility of Bitcoin Returns

Wonse Kim\*, Junseok Lee†, Kyungwon Kang#

**Abstract**

This paper investigates the effects of the launch of Bitcoin futures on the intraday volatility of Bitcoin. Based on one-minute price data collected from four cryptocurrency exchanges, we first examine the change in *realized volatility* after the introduction of Bitcoin futures to investigate their aggregate effects on the intraday volatility of Bitcoin. We then analyze the effects in more detail utilizing the discrete Fourier transform. We show that although the Bitcoin market became more volatile immediately after the introduction of Bitcoin futures, over time it has become more stable than it was before the introduction.

_______________________________________

\* Corresponding author: Research institute of Mathematics, Seoul National University, Seoul, Korea. E-mail: aquinasws@snu.ac.kr, Tel.: +82-10-3071-7851.

†: Graduate School, Department of Mathematics, Seoul National University, Seoul, Korea. E-mail: asd1206@snu.ac.kr, Tel.: +82-10-4139-3307.

#Team Presto, Sharon Bld., Nambusunhwan-ro, Gwanak-gu, Seoul, Korea E-mail: presto@presto-platform.io C/P: +82-10-9371-7940

## 1. Introduction

With the recent substantial increase in attention to the Bitcoin market (Urquhart 2016, Katsiampa 2017), a growing number of studies have investigated the that market's properties.[1] In particular, since the Chicago Mercantile Exchange (CME) and the Chicago Board Options Exchange (CBOE) launched Bitcoin futures in December 2017,[2] several studies have sought to uncover the effect of that introduction on the Bitcoin market. Hale et al. (2018) suggested that, by allowing pessimists to enter the market, the launch of Bitcoin futures contributed to a rapid decline in Bitcoin's value immediately after its appearance. Employing one-minute data from September 26, 2017, to February 22, 2018, Corbet et al. (2018) showed that volatility increased around the announcement of Bitcoin futures.

However, because this line of research investigates the relatively short-term (usually less than three months) effect of the introduction of Bitcoin futures on the Bitcoin market, the question concerning the long-term effect remains. This article fills in this research gap by investigating the long-term effect of the launch of Bitcoin futures on the Bitcoin market's stability. As in the case of stock market, there might be two conflicting explanations for the effect of Bitcoin futures on the Bitcoin market's stability (Lee and Ohk 1992): One is that, as Freidman (1953) noted, since uninformed speculators will be eliminated quickly from the market, the trades of Bitcoin futures by well-informed speculators stabilize the market. The other is that, as trading volumes are diverted from the underlying Bitcoin to Bitcoin futures, the decreased liquidity in the underlying Bitcoin may increase the futures' volatility. In this article, we utilize Kim and Jun's (2018) approach and one-minute Bitcoin price data from the five largest cryptocurrency exchanges during the six months before and after the introduction of the Bitcoin futures to examine these competing explanations.

Our approach consists of two steps. First, to investigate the aggregate change in Bitcoin's intraday volatility, we examine the change in the bias-corrected *realized volatility* (Hansen, Lunde, 2006) after Bitcoin futures were introduced. Then we use the discrete Fourier transform (DFT) to analyze the change in more detail.

DFT provides additional advantages in studying intraday volatility over using the realized variance alone: First, whereas traditional intraday volatility measures usually assume a specific parametric stochastic process model as the price process[3], DFT is fully model-free and non-parametric. Hence, the analysis based on DFT provides more realistic results than those from an analysis based on traditional volatility measures. In addition, as Malliavin and Mancino (2002) note, since DFT's algorithm is based on integrating time series, DFT is more robust than traditional methods are.[4]

By performing an analysis based on DFT in addition to an analysis using realized volatility, we can get a realistic, robust, and detailed results on the change in intraday Bitcoin volatility after trading of Bitcoin futures was introduced (Kim and Jun 2018). To the best of our knowledge, this study is the first to investigate empirically the long-term effect of the launch of Bitcoin futures on Bitcoin's market stability.

## 2. Data

Although most empirical studies on cryptocurrency have used data from one exchange, to ensure the validity of our empirical results we use data from the five largest cryptocurrency exchanges: Bitflyer, Coincheck (Japan), Bitstamp (E.U), Coinbase (U.S) and Binance (Hong Kong). In addition, our sample period includes an additional four months of data before and after the sample period used in Corbet et al. (2018) to allow us to investigate the long-term effects of introducing Bitcoin futures. The dataset from Coinbase and Bitstamp (Coincheck) consists of one-minute prices for Bitcoin in USD (JPY) from June 1, 2017, to June 26, 2018; the dataset from Bitflyer consists of one-minute prices for Bitcoin in JPY from July 5, 2017, to June 26, 2018; and the dataset from Binance consists of one-minute prices for Bitcoin and Ethereum from September 4, 2017, to June 5, 2018. We split each dataset into daily categories based on the New York time zone (Eastern Time).[5]

---

[3] For example, the realized volatility assumes that price processes follow geometric Brownian motion.
[4] According to Malliavin and Mancino (2002), most traditional volatility measures use an algorithm based on a "differentiation procedure" that is highly unstable.
[5] Our empirical results are robust for other time zones, such as London and Tokyo.

Since Bitcoin futures began trading on CME and CBOE in December 2017, our sample period covers about six months before and six months after the introduction of the futures. To investigate the long-term effect of introducing Bitcoin futures, we split our dataset into four sub-periods: *Period 0* covers the period before the introduction of the Bitcoin futures; *Period 1* runs from December 18[6], 2017, to February 28, 2018; *Period 2* runs from March 1, 2018, to April 30, 2018; and *Period 3* is from May 1, 2018, to June 26, 2018.[7] Periods 1, 2, and 3 are split so each sub period covers an even two months.[8] Figure 1 shows Bitcoin's daily closing prices during our sample period.

<center>Insert Figure 1 about here.</center>

### 3. Methodology and Results

To assess the *aggregate* change in the intraday volatility of Bitcoin's price process after the introduction of Bitcoin futures, we estimated the bias-corrected *realized volatility* (Hansen and Lunde, 2006). The bias-corrected realized volatility of Bitcoin on day $t$, $\sigma_t$, is defined as:

$$\sigma_t \; = \; \sqrt{\sum_{k=1}^{1440} r_{t,k}^2 + 2\frac{1440}{1339}\sum_{k=1}^{1339} r_{t,k} r_{t+1,k}} \; , \qquad (1)$$

where $r_{t,k}$ is the $k$-th one-minute log return of Bitcoin on day $t$. Since Bitcoin is traded for twenty-four hours, there are 1,440 returns for each day.

Table 1 reports the time series averages of Bitcoin's bias-corrected realized volatility with their corresponding standard errors for each exchange during the four periods. Table 1 shows that Bitcoin's realized volatility statistically significantly increased in Period 1 compared to Period 0, which is consistent with Corbet et al. (2018). However, the realized volatility gradually decreased in Periods 2 and 3, and Period 3's realized volatility decreased statistically significantly compared to Period 0. Therefore, the results shown in Table 1 suggest that, although Bitcoin's realized volatility increased immediately after the futures were introduced, the realized volatility decreased gradually after that to a level lower than it was before the futures were introduced. This finding holds true for all five exchanges in our dataset. Table 1 also contains the empirical results from the same analysis for the *Garman-Klass volatility estimator*[9] to gauge the robustness of our result and shows that our main results are robust for that estimator.

---

[6] Although CBOE opened a futures market on December 10, trading volume was too small until the CME launched Bitcoin futures (Hale et al., 2018) on December 18, 2017, so we choose December 18, 2017, as the day when Bitcoin futures were introduced. Our empirical results remained unchanged even when we used December 10, 2017, as the day of the introduction.

[7] For Binance, *Period 3* is from May 1, 2018, to June 5, 2018.

[8] Our empirical results are consistent with a diverse selection of sub-periods.

<p align="center">Insert Table 1 about here.</p>

To estimate the causal effects of the introduction of Bitcoin futures more precisely, we perform the difference-in-differences (DD) analysis. We use Bitcoin prices from the Binance exchange as a treatment variable and the Ethereum price from the same exchange as a control variable. Using Bitcoin and Ethereum price data from the Binance exchange, we estimate the following regression equation for the DD estimator for a fixed *k (k = 1, 2, or 3)*:

$$\log(\sigma_t) = \alpha + \beta_1 (Period[k]) + \beta_2 (Treatment) + \beta_3 (\text{Period} \times Treatment) + \epsilon, \quad (2)$$

where $Period[k]$ is a time dummy variable that equals 1 for observations obtained in Period $k$, the time after the introduction of Bitcoin futures, and zero for observations in Period_0, the time before the introduction of Bitcoin futures. $Treatment$ is a dummy variable that switches on for observations of Bitcoin and off for observations of Ethereum.

<p align="center">Insert Table 2 about here.</p>

Table 2, which presents the coefficient estimates for the regression model (2) with corresponding t-statistics, shows that, although negative estimates of $\beta_3$ for $k = 1$ and $k = 2$ are statistically insignificant, the negative estimate of $\beta_3$ for k = 3 is statistically significant. Thus, the empirical results shown in Table 2 suggest that, although the increased (decreased) intraday volatility

---

[9] The Garman-Klass volatility estimator is defined by

$$\sigma_t^{GK} = \sqrt{\sum_{k=1}^{1440} \frac{1}{2} \left( \log \left( \frac{H_k}{L_k} \right) \right)^2 - (2 \cdot \log(2) - 1) \cdot \left( \log \left( \frac{C_k}{O_k} \right) \right)^2} \,,$$

where $O_k$, $H_k$, $L_k$, and $C_k$ are the open, high, low, and close prices, respectively, during the *k*-th one minute .

of Bitcoin in Period 1 (Period 2) compared to Period 0 may not have been caused by Bitcoin futures, the decreased intraday volatility of Bitcoin in Period 3 was. Therefore, the results from Table 2 suggest that, in the long run, the Bitcoin futures contribute to making the market more stabilized than it was before the futures were introduced.

To analyze the change in the intraday volatility of Bitcoin prices in more detail, we applied DFT to the intraday Bitcoin price time series, as in Kim and Jun (2018). The Fourier coefficients of Bitcoin's price time series on day $t$ are given by:

$$a_t(w) = \frac{2}{1441} \sum_{k=1}^{1441} P_t(k) \cdot cos(2\pi wk/1441) \qquad (3)$$

$$b_t(w) = \frac{2}{1441} \sum_{k=1}^{1441} P_t(k) \cdot sin(2\pi wk/1441) \qquad (4)$$

for $w = 1, 2, \ldots, 720$, , where $P_t(k)$ denotes the $k$-th one-minute log price of Bitcoin on day $t$. Thus, we obtain a 720-element amplitude vector of Bitcoin price on day $t$,[10] $C_t$. The $w$-th element of the vector $C_t$ represents the amplitude of frequency $w$, such that $C_t(w)$ $\left( = \sqrt{a_t(w)^2 + b_t(w)^2} \right)$.

Using computed $C_t$ for each day $t$, we calculated the time series average amplitude of each frequency component for each sub-period:

$$\overline{C(w)}_{Period\_i} = \sqrt{\frac{\sum_{t \in Period\_i} C_t(w)^2}{|Period\_i|}} \quad , \qquad (5)$$

where *Period_i* is the set of days in our sample period that belong to Period *i*.

Insert Figure 2 about here.

Figure 2 shows the change rates of the time series averages of $w$-frequency components *Change_Freq$_w$* for Periods 1, 2, and 3 as compared to Period 0 for the five cryptocurrency exchanges. Figure 2 shows that the *Change_Freq$_w$* for Period 2 is around 1 and that the *Change_Freq$_w$* of every frequency component for Period 3 is smaller than 1, whereas the *Change_Freq$_w$* for every frequency component in Period 1 is larger than 1. To provide more statistical analysis from the empirical results from DFT, we use Hau's (2001) approach, which groups frequen Insert Figure 2 about here cies into three frequency bands: a low-frequency band, a medium-frequency band, and a high-frequency band. Table 3 reports the average $Change\_Freq_w$ and t-statistics for the one sample t test determining whether the average is statistically different from 1, for the low-frequency band ($w$: 1-240) , the medium-frequency band ($w$: 241-480) , and the high-frequency band ($w$: 481-720) for Periods 1, 2, and 3, respectively over Period 0, and for the five cryptocurrency exchanges. Table 3 shows that, for all five exchanges, the average $Change\_Freq_w$ of low frequency is statistically larger than 1, whereas those of medium- and high frequency are statistically smaller than 1. Moreover, the average $Change\_Freq_w$ of the low-, medium-, and high-frequency component of Period 3 are statistically significantly smaller than those of Period 2. The empirical results in Table 3 indicate that all of the low-, medium-, and high-frequency components of the Bitcoin price time series increased in Period 1 compared to Period 0 and that all of them significantly decreased in Periods 2 and 3 compared to Period 0. In addition, the low-, medium-, and high-frequency components of the Bitcoin price time series decreased in Period 3 compared to Period 2. Since the sample variance of a stochastic process can be expressed as the sum of squares of the sample's amplitudes for all frequencies of the stochastic process (see Hamilton, 1994, section 6.2), the results in Table 3 suggest that it is not the change of partial frequency components but the changes in the low-, medium-, and high-frequency components that induce the intraday volatility changes following the appearance of the Bitcoin futures reported in Table 1.

---

[10] Since the Bitcoin market is open for twenty-four hours, we obtained a 1,441-dimension price vector array.

Using two–regime Markov–switching GARCH models and a dataset that ended on March 3, 2018, Ardia et al. (2018) also showed that there had been a switch to the high-volatility regime around the launch of the Bitcoin futures. Since their dataset does not cover Periods 2 and 3, to show that their preferred model specification provides the same result that our model does during Periods 2 and 3, we investigate the regime change in the GARCH volatility dynamics of Bitcoin log-returns using the two-regime skewed Student–t GJR model, which is the preferred model in Ardia et al. (2018).[11]

---

[11] Because the number of observations in our sample is not high enough to employ MCMC simulation, we performed Maximum Likelihood estimation using the R package developed by Ardia et al. (2016).

Figure 3 shows the smoothed probabilities for the two-regime skewed Student–t GJR model during our sample period. The figure shows that, whereas the probability of a high-volatility regime is near 1 during Periods 0 and 1, the probability of a high-volatility regime starts to approach 0 and the probability of a low-volatility regime starts to approach 1 in the middle of Period 2 (the beginning of April, 2018). Therefore, the empirical result from the two-regime Markov-switching GARCH model analysis also supports our main finding that, in the long-run, the Bitcoin price process was more stable than its previous level following the appearance of Bitcoin futures.

Insert Figure 3 about here

## 4. Conclusions

We used the bias-corrected realized volatility and the DFT to investigate the effects of Bitcoin futures' introduction on the intraday volatility of Bitcoin based on one-minute Bitcoin price data. Our study has two primary findings: First, for a short time immediately after the trading of Bitcoin futures began, the realized volatility increased and the low-, medium, and high-frequency components of the Bitcoin price process increased to points higher than they were before the futures were introduced. This result suggests that the Bitcoin market was not favorable to liquidity-providers for that short period after the introduction of Bitcoin futures because they risked increased low-frequency price swings, even with the same inventories. Second, as time passed, both the realized volatility and all of the frequencies of the Bitcoin price process decreased to below where they were before Bitcoin futures were introduced. These two findings show that, although the Bitcoin market became unstable for a while immediately after the introduction of the futures market, over time the market gradually became more stabilized than it was before.

**Acknowledgment**

This research was supported by BK21 PLUS SNU Mathematical Sciences Division.

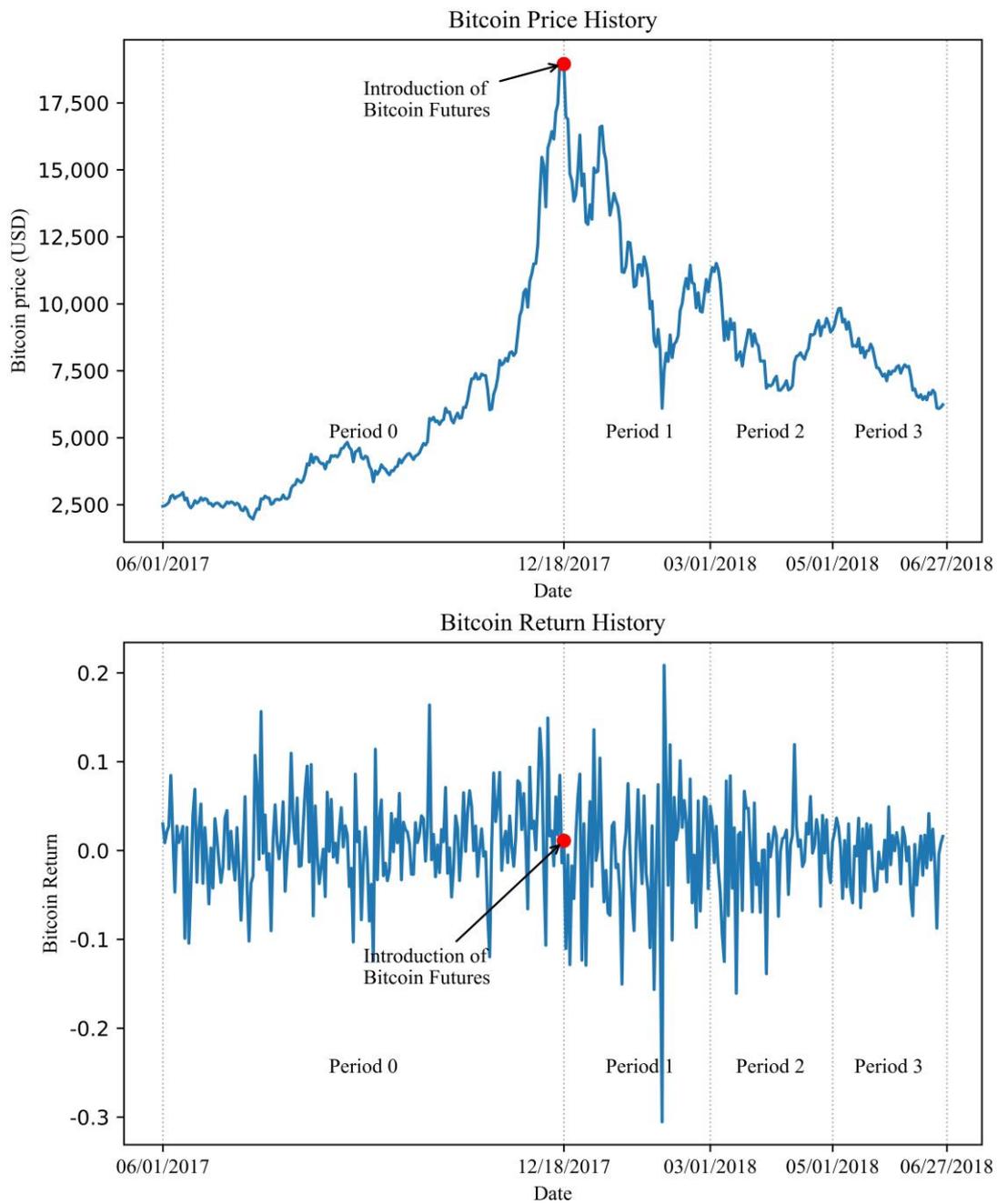

Fig. 1 Time-series graph of Bitcoin's daily closing price and daily log return from June 1, 2017, to June 26, 2018.

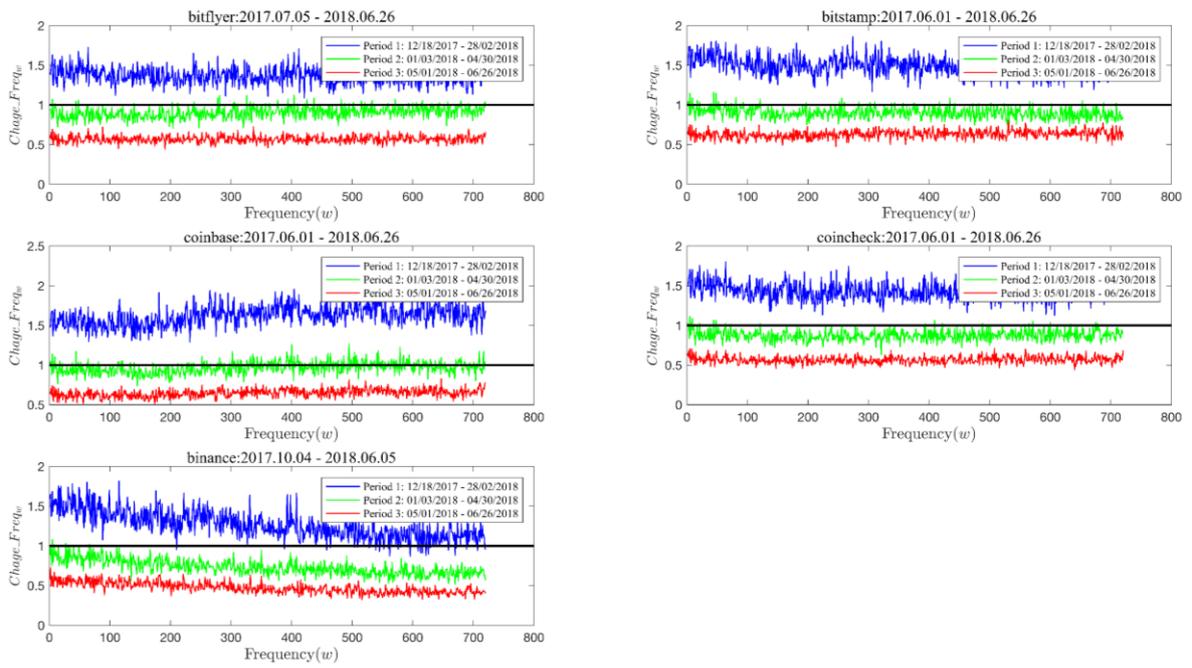

Fig. 2 The change rates of the time series averages of *w*-frequency components, *Change_Freq*$_w$ for

Periods 1, 2, and 3 over Period 0 for Bitflyer, Coincheck, Bitstamp, Coinbase, and Binance exchanges.

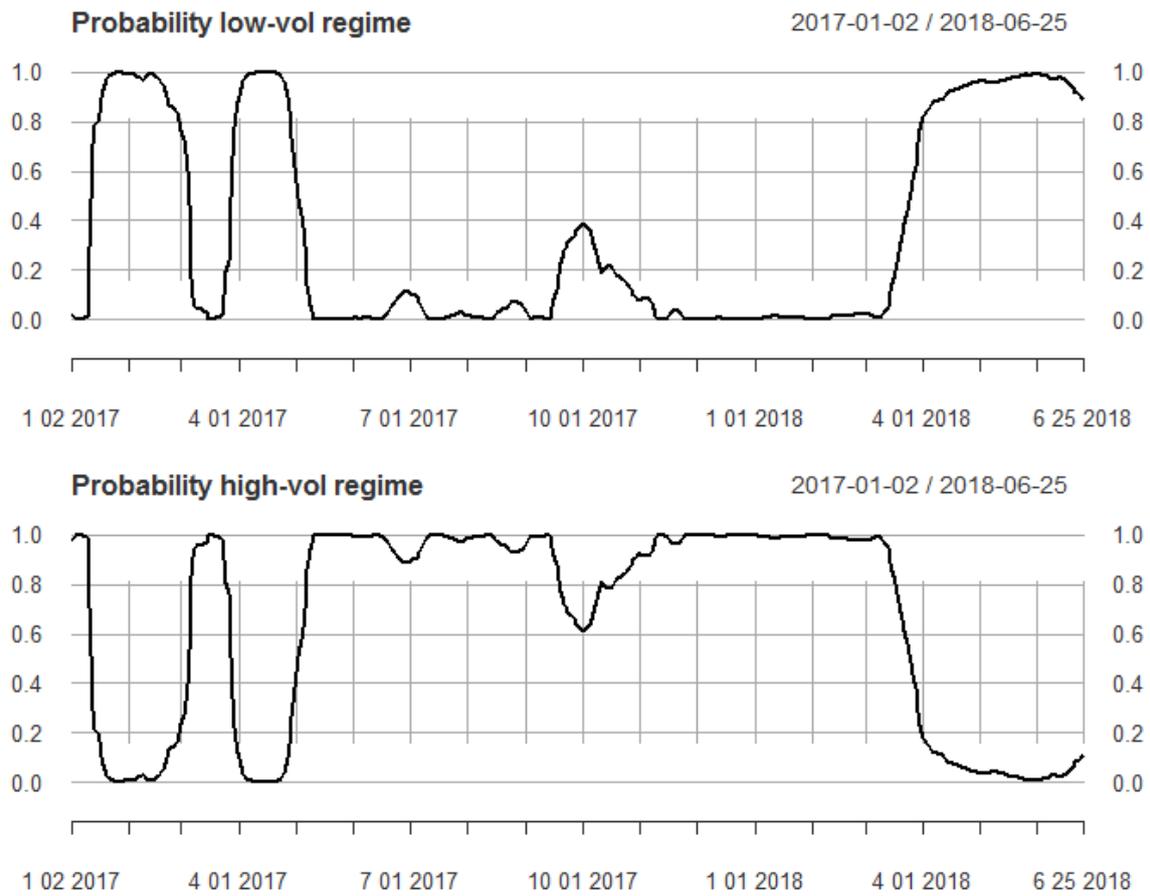

Fig. 3 Smoothed probabilities for the two–regime skewed Student–t GJR model of Bitcoin returns.

Table 1. Time series averages of the realized volatility of Bitcoin prices for four periods. Standard errors are reported in brackets, and t-statistics are in parentheses.

**Panel A. Bitflyer**

| Period | Period 0: 07/05/2017 - 12/17/2017 | Period 1: 12/18/2017 - 02/28/2018 | Period 2: 03/01/2018 - 04/30/2018 | Period 3: 05/01/2018 - 06/26/2018 |
|---|---|---|---|---|
| $\overline{\sigma}_{Period\_i}$ ($i$=0,1,2,3) | 0.0502 [0.0025] | 0.0721 [0.0041] | 0.0480 [0.0031] | 0.0295 [0.0011] |
| $\overline{\sigma^{GK}}_{Period\_i}$ ($i$=0,1,2,3) | 0.0502 [0.0025] | 0.0721 [0.0041] | 0.0480 [0.0031] | 0.0295 [0.0011] |
| $\overline{\sigma}_{Period\_i} - \overline{\sigma}_{Period\_0}$ ($i$=1,2,3) | | -0.0215 (-4.5948) | 0.0026 (0.5665) | 0.0207 (4.7218) |
| $\overline{\sigma^{GK}}_{Period\_i} - \overline{\sigma^{GK}}_{Period\_0}$ ($i$=1,2,3) | | -0.0215 (-4.5948) | 0.0026 (0.5665) | 0.0207 (4.7218) |

**Panel B. Bitstamp**

| Period | Period 0: 07/05/2017 - 12/17/2017 | Period 1: 12/18/2017 - 02/28/2018 | Period 2: 03/01/2018 - 04/30/2018 | Period 3: 05/01/2018 - 06/26/2018 |
|---|---|---|---|---|
| $\overline{\sigma}_{Period\_i}$ ($i$=0,1,2,3) | 0.0502 [0.0025] | 0.0721 [0.0041] | 0.0480 [0.0031] | 0.0295 [0.0011] |
| $\overline{\sigma^{GK}}_{Period\_i}$ ($i$=0,1,2,3) | 0.0502 [0.0025] | 0.0721 [0.0041] | 0.0480 [0.0031] | 0.0295 [0.0011] |
| $\overline{\sigma}_{Period\_i} - \overline{\sigma}_{Period\_0}$ ($i$=1,2,3) | | -0.0215 (-4.5948) | 0.0026 (0.5665) | 0.0207 (4.7218) |
| $\overline{\sigma^{GK}}_{Period\_i} - \overline{\sigma^{GK}}_{Period\_0}$ ($i$=1,2,3) | | -0.0215 (-4.5948) | 0.0026 (0.5665) | 0.0207 (4.7218) |

**Panel C. Coinbase**

| Period | Period 0: 07/05/2017 - 12/17/2017 | Period 1: 12/18/2017 - 02/28/2018 | Period 2: 03/01/2018 - 04/30/2018 | Period 3: 05/01/2018 - 06/26/2018 |
|---|---|---|---|---|
| $\overline{\sigma}_{Period\_i}$ ($i$=0,1,2,3) | 0.0502 [0.0025] | 0.0721 [0.0041] | 0.0480 [0.0031] | 0.0295 [0.0011] |
| $\overline{\sigma^{GK}}_{Period\_i}$ ($i$=0,1,2,3) | 0.0502 [0.0025] | 0.0721 [0.0041] | 0.0480 [0.0031] | 0.0295 [0.0011] |
| $\overline{\sigma}_{Period\_i} - \overline{\sigma}_{Period\_0}$ ($i$=1,2,3) | | -0.0215 (-4.5948) | 0.0026 (0.5665) | 0.0207 (4.7218) |
| $\overline{\sigma^{GK}}_{Period\_i} - \overline{\sigma^{GK}}_{Period\_0}$ ($i$=1,2,3) | | -0.0215 (-4.5948) | 0.0026 (0.5665) | 0.0207 (4.7218) |

**Panel D. Coincheck**

| Period | Period 0: 07/05/2017 - 12/17/2017 | Period 1: 12/18/2017 - 02/28/2018 | Period 2: 03/01/2018 - 04/30/2018 | Period 3: 05/01/2018 - 06/26/2018 |
|---|---|---|---|---|
| $\overline{\sigma}_{Period\_i}$ ($i$=0,1,2,3) | 0.0502 [0.0025] | 0.0721 [0.0041] | 0.0480 [0.0031] | 0.0295 [0.0011] |
| $\overline{\sigma^{GK}}_{Period\_i}$ ($i$=0,1,2,3) | 0.0502 [0.0025] | 0.0721 [0.0041] | 0.0480 [0.0031] | 0.0295 [0.0011] |
| $\overline{\sigma}_{Period\_i} - \overline{\sigma}_{Period\_0}$ ($i$=1,2,3) | | -0.0215 (-4.5948) | 0.0026 (0.5665) | 0.0207 (4.7218) |
| $\overline{\sigma^{GK}}_{Period\_i} - \overline{\sigma^{GK}}_{Period\_0}$ $(i=1,2,3)$ | | -0.0215 (-4.5948) | 0.0026 (0.5665) | 0.0207 (4.7218) |

Panel E. Binance

| Period | Period 0: 07/05/2017 - 12/17/2017 | Period 1: 12/18/2017 - 02/28/2018 | Period 2: 03/01/2018 - 04/30/2018 | Period 3: 05/01/2018 - 06/26/2018 |
|---|---|---|---|---|
| $\overline{\sigma}_{Period\_i}$ $(i=0,1,2,3)$ | 0.0502 [0.0025] | 0.0721 [0.0041] | 0.0480 [0.0031] | 0.0295 [0.0011] |
| $\overline{\sigma^{GK}}_{Period\_i}$ $(i=0,1,2,3)$ | 0.0502 [0.0025] | 0.0721 [0.0041] | 0.0480 [0.0031] | 0.0295 [0.0011] |
| | | | | |
| $\overline{\sigma}_{Period\_i} - \overline{\sigma}_{Period\_0}$ $(i=1,2,3)$ | | -0.0215 (-4.5948) | 0.0026 (0.5665) | 0.0207 (4.7218) |
| $\overline{\sigma^{GK}}_{Period\_i} - \overline{\sigma^{GK}}_{Period\_0}$ $(i=1,2,3)$ | | -0.0215 (-4.5948) | 0.0026 (0.5665) | 0.0207 (4.7218) |

Table 2 coefficient estimates for the regression model:

$$\log(\sigma_t) = \alpha + \beta_1(Period[k]) + \beta_2(Treatment) + \beta_3(Period \times Treatment) + \epsilon$$

t-statistics are in parentheses.

| coefficients | $k$ | | |
|---|---|---|---|
| | 1 | 2 | 3 |
| $\alpha$ | -2.8504 | -2.8504 | -2.8504 |
| | (-54.8690) | (-64.1285) | (-66.4862) |
| $\beta_1$ | 0.3755 | -0.1466 | -0.2685 |
| | (4.9016) | (-2.2069) | (-3.5493) |
| $\beta_2$ | -0.0357 | -0.0357 | -0.0357 |
| | (-0.4861) | (-0.5681) | (-0.5890) |
| $\beta_3$ | 0.0507 | -0.0242 | -0.2872 |
| | (0.4677) | (-0.2578) | (-2.6845) |

Table 3. The averages of $Change\_Freq_w$ for the low-frequency band ($w: 1 - 240$), the medium-frequency band ($w: 241 - 480$), and the high-frequency band ($w: 481 - 720$). t-statistics for the one sample t test determining whether the average is statistically different from 1 are in parentheses.

|  | Low-frequency band (frequency: 1 -240) | Medium-frequency band (frequency: 241 -480) | High-frequency band (frequency: 481 -720) |
|---|---|---|---|
| | Panel A. Bitflyer | | |
| Period_1 over Period_0 | 1.3790 | 1.3632 | 1.3517 |
| | (56.8441) | (53.4469) | (55.0310) |
| Period_2 over Period_0 | 0.8710 | 0.9069 | 0.9303 |
| | (-29.4229) | (-19.7634) | (-17.9206) |
| Period_3 over Period_0 | 0.5664 | 0.5658 | 0.5716 |
| | (-149.9616) | (-158.1799) | (-151.0844) |
| | Panel B. Coincheck | | |
| Period_1 over Period_0 | 1.4462 | 1.4161 | 1.4275 |
| | (58.5555) | (62.8475) | (63.3660) |
| Period_2 over Period_0 | 0.8732 | 0.8650 | 0.8841 |
| | (-28.1780) | (-32.9304) | (-28.5885) |
| Period_3 over Period_0 | 0.5647 | 0.5587 | 0.5754 |
| | (-149.9566) | (-170.1349) | (-142.0954) |
| | Panel C. Bitstamp | | |
| Period_1 over Period_0 | 1.5207 | 1.4982 | 1.4348 |
| | (68.3596) | (67.0054) | (67.1433) |
| Period_2 over Period_0 | 0.9123 | 0.9019 | 0.8824 |
| | (-19.0524) | (-23.5343) | (-26.8819) |
| Period_3 over Period_0 | 0.6126 | 0.6343 | 0.6435 |
| | (-121.5173) | (-116.2992) | (-110.7876) |
| | Panel D. Coinbase | | |
| Period_1 over Period_0 | 1.5352 | 1.6607 | 1.6712 |
| | (73.5442) | (91.9911) | (87.5302) |
| Period_2 over Period_0 | 0.9254 | 0.9787 | 0.9926 |
| | (-16.9338) | (-4.3252) | (-1.5977) |
| Period_3 over Period_0 | 0.6208 | 0.6534 | 0.6690 |
| | (-122.1348) | (-115.6467) | (-102.4669) |
| Panel E. Binance | | | |
| Period_1 over Period_0 | 1.4109 | 1.2578 | 1.1284 |
| | (40.9436) | (30.5650) | (18.1477) |
| Period_2 over Period_0 | 0.8134 | 0.7182 | 0.6621 |
| | (-33.2399) | (-66.5696) | (-84.7175) |
| Period_3 over Period_0 | 0.5318 | 0.4621 | 0.4212 |
| | (-127.1549) | (-164.8501) | (-208.2267) |

Appendix

Panel A.
Some facts based on CBOE and CME Bitcoin futures.

|  | CBOE | CME |
|---|---|---|
| Product Code | XBT | BTC |
| Listing Date | December 10, 2017 | December 18, 2017 |
| Contract Unit | 1 Bitcoin | 5 Bitcoins |
| Trading hours | 9:30 a.m. - 4:15 p.m. ET Monday – Friday | 6:00 p.m. - 5:00 p.m. ET Monday – Friday |
| Settlement | The final settlement value of an expiring XBT futures contract shall be the official auction price for bitcoin in U.S. dollars determined at 4:00 p.m. Eastern Time on the final settlement date by the Gemini Exchange (the "Gemini Exchange Auction") | Cash settled by reference to the Final Settlement Price. |

Panel B.
Daily descriptive statistics on CBOE's Bitcoin futures.

| Future | Trade Volume | Open Interest | Daily Volatility |
|---|---|---|---|
| CFE_F18_XBT | 3990.2 | 2413.8 | 0.005834 |
| CFE_F19_XBT | 270.5 | 623.4 | 0.002397 |
| CFE_G18_XBT | 3568.9 | 1926.9 | 0.005484 |
| CFE_G19_XBT | 45.8 | 269.0 | 0.003369 |
| CFE_H18_XBT | 2108.7 | 2193.1 | 0.004868 |
| CFE_H19_XBT | 49.4 | 195.6 | 0.002321 |
| CFE_J18_XBT | 2350.3 | 2122.4 | 0.003601 |
| CFE_K18_XBT | 1725.2 | 1532.5 | 0.002954 |
| CFE_M18_XBT | 1197.7 | 1489.7 | 0.001963 |
| CFE_N18_XBT | 998.5 | 1389.4 | 0.001775 |
| CFE_Q18_XBT | 1382.6 | 957.2 | 0.001699 |
| CFE_U18_XBT | 972.6 | 1273.6 | 0.001652 |
| CFE_V18_XBT | 616.2 | 1010.7 | 0.001194 |
| CFE_X18_XBT | 365.6 | 724.0 | 0.000991 |
| CFE_Z18_XBT | 1334.4 | 1299.9 | 0.001857 |